\documentclass[12pt,journal,onecolumn]{IEEEtran}

\usepackage{cite,authblk,rotating,setspace,amsmath,amssymb,bbm, theorem,epstopdf}
\usepackage[caption=false,labelformat=simple, font=footnotesize]{subfig}

\usepackage{tabularx}
\usepackage{adjustbox}
\usepackage{booktabs}
\usepackage{multirow}

\IEEEoverridecommandlockouts
\allowdisplaybreaks

\begin{document}

        \title{Multivariate, Multi-step, and Spatiotemporal Traffic Prediction for NextG Network Slicing under SLA Constraints}

        \author[1,2]{Evren Tuna}
        \author[2,3]{Alkan Soysal}
        \affil[1]{\normalsize Department of Research \& Development, ULAK Communications Inc., Ankara, Turkey}
        \affil[2]{\normalsize Department of Electrical and Electronics Engineering, Bahcesehir University, Istanbul, Turkey}
        \affil[3]{\normalsize Department of Electrical and Computer Engineering, Virginia Tech, Blacksburg, VA 24061}

        \maketitle
\vspace{-0.5in}

\begin{abstract}

This study presents a spatiotemporal traffic prediction approach for NextG mobile networks, ensuring the service-level agreements (SLAs) of each network slice. Our approach is multivariate, multi-step, and spatiotemporal. Leveraging 20 radio access network (RAN) features, peak traffic hour data, and mobility-based clustering, we propose a parametric SLA-based loss function to guarantee an SLA violation rate. We focus on single-cell, multi-cell, and slice-based prediction approaches and present a detailed comparative analysis of their performances, strengths, and limitations. 

First, we address the application of single-cell and multi-cell training architectures. While single-cell training offers individual cell-level prediction, multi-cell training involves training a model using traffic from multiple cells from the same or different base stations. We show that the single-cell approach outperforms the multi-cell approach and results in test loss improvements of 11.4\% and 38.1\% compared to baseline SLA-based and MAE-based models, respectively.

Next, we explore slice-based traffic prediction. We present single-slice and multi-slice methods for slice-based downlink traffic volume prediction, arguing that multi-slice prediction offers a more accurate forecast. The slice-based model we introduce offers substantial test loss improvements of 28.2\%, 36.4\%, and 55.6\% compared to our cell-based model, the baseline SLA-based model, and the baseline MAE-based model, respectively.

\end{abstract}
%


%
%
\section{Introduction}

The integration of key enablers, such as network slicing, private networks, and edge computing, tailored to specific services, has escalated the complexity of Next Generation (NextG) networks. Intelligent network management based on Machine Learning (ML) can address this complexity, enhancing the robustness, predictiveness, autonomy, and reliability of NextG cellular networks \cite{Shafin2020, Zhang2019, Wang2018}. For instance, the Open Radio Access Network (RAN) paradigm employs near-real-time (10-1000 ms) and non-real-time (greater than 1 s) RAN Intelligent Controllers (RIC) for network management and control \cite{Polese2023}. A network traffic forecast application at the non-real-time RIC can support applications at both the near-real-time RIC, such as network slicing, scheduling, mobility management, and radio resource management, and at the non-real-time RIC, like energy efficiency management.

Predicting network traffic in mobile networks necessitates analysis of substantial, highly dynamic past data, representing a complex, multi-dimensional problem. A comprehensive understanding of this problem from both data and communications perspectives is pivotal for implementing a practical solution in Open RAN. Notable dimensions in a NextG traffic prediction problem include temporal effects, spatial granularity, spatial clustering, feature selection, problem-specific optimization objectives, prediction horizon, and the transferability of the trained model to the operator's entire system.

Traditionally, temporal effects in time-series problems were addressed by variations of Autoregressive Integrated Moving Average (ARIMA) models \cite{Bui2017}. For instance, \cite{Xu2016} divided time series mobile traffic volume data into regular and random components. The regular component was predicted using ARIMA, resulting in a 30\% error, whereas the random component was deemed unpredictable. Recently, ML methods have been applied to traffic volume prediction, yielding significantly better results than ARIMA models. Some ML-based works considered only temporal effects without spatial analysis where Long Short Term Memory (LSTM)-based Recurrent Neural Network (RNN) models outperformed statistical models like ARIMA \cite{Trinh2018, Gutterman2019, Wang2020b, Chergui2020}. These temporal-only analyses primarily focused on single time-series data from a single cell or aggregated data from multiple cells. 

While prior works do not differentiate between spatial granularity and spatial clustering, it is essential to understand the distinct impacts of both dimensions. When working with multiple time-series data from multiple cells, most prior works employ a grid-based spatial granularity, dividing the entire service area into smaller grids \cite{Wang2017, Huang2017, Chen2018a, Zhang2018, Zhang2018b, Zhang2019c}. Traffic data from cells within the same grid are aggregated and treated as a single time series. This technique possibly results in lower prediction errors due to data smoothing but hinders mapping predicted data back to individual cells for cell-level intelligent control. The majority of grid-based approaches use Convolutional Neural Network (CNN)-based models to leverage spatial dependencies, drawing from CNN structures used in video processing applications. The matrix-format inputs represent the aggregated traffic of base stations in corresponding square grid areas. However, a persistent issue with grid-based spatial precision is its limited generalizability to a provider's entire service area. Despite these challenges, the grid-based approach has been gaining traction \cite{Li2020, Zhao2020b, Shen2021, Liu2021b, Wu2022c, Li2022d, Hu2023, Rao2023, Wang2023b, Yao2023, Mehrabian2023} leading to increasingly complex models that are likely overfitting and challenging to implement at the non-real-time RIC level.

In contrast to grid-level spatial granularity, other works have examined base station level \cite{Qiu2018, Feng2018, Gao2019, Xing2021, Bega2020, Wang2022b, Chen2023, Wang2019} and cell level \cite{Zhao2020, Fang2018, Tuna2022, Tuna2023} spatial granularity. References \cite{Bega2020, Wang2022b} use CNN at the base station level by rearranging base stations into a matrix structure. Conversely, \cite{Fang2018, Wang2019, Zhao2020, Chen2023} propose using graph convolutional networks (GCNs) instead. Here, the base stations or the cells are the graph nodes, and edges represent some interaction between cells. GCN implementation is challenging due to the need for pre-existing knowledge of all cell adjacency matrices, which should remain static over time. As a result, GCN may not be feasible for large-scale wireless systems \cite{Mehrabian2023}.

The same works can also be classified according to their spatial clustering approach. Clustering is beneficial when data points (grids, base stations, cells) in the cluster interact. In proximity-based clustering, data points in close proximity are processed together \cite{Huang2017, Chen2018a, Zhang2018, Zhang2018b, Li2020, Zhao2020b, Shen2021, Liu2021b, Wu2022c, Li2022d, Hu2023, Rao2023, Qiu2018, Feng2018, Gao2019, Fang2018}. However, proximity does not necessarily indicate an interaction between cells \cite{Wang2019}. In similarity-based clustering, data points exhibiting a certain similarity measure are processed together. Multiple similarity measures are considered in the literature, including correlation, k-shape, dynamic time warping, and spectral decomposition. Similarity does not necessarily indicate an interaction between cells, either. Alternatively, data volume at a neighboring cell might affect the data volume at the intended cell if there is handover between the cells. In mobility-based clustering, handover data is used to cluster data points \cite{Wang2019, Zhao2020, Tuna2022, Tuna2023}. While \cite{Wang2019, Zhao2020} use handover data to calculate edge values in a GCN and suffer from GCN's shortcomings, our previous work \cite{Tuna2022, Tuna2023} can be dynamically applied to an entire network. Building upon \cite{Tuna2022, Tuna2023}, we expand the scope of the research in this paper to incorporate multi-cell and slice-based approaches. 

The majority of previous works utilize standard loss functions such as MAE, MSE, and their variants to minimize prediction errors. These symmetrical loss functions treat SLA violations and overprovisioning the same, resulting in nearly identical SLA violation and overprovisioning rates. However, operators prefer overprovisioning to SLA violations due to stringent SLA obligations, especially for slice-based network traffic. Since it is crucial to choose problem-specific optimization objectives, we propose a Service-Level Agreements (SLA) violation-based loss function for the traffic prediction problem in NextG network slicing. Traditional loss functions, such as MSE and MAE, are unsuitable for mobile operators who must maximize resource utilization while avoiding any SLA violations on network slices. Despite its importance, literature is limited in  applications of SLA-based loss functions \cite{Bega2020, Tuna2022, Tuna2023}. 

Our approach in this study is multivariate, multi-step, spatiotemporal, and SLA-driven. We investigate the effect of 20 different RAN features on predicting future values of downlink traffic volume. We devise additional feature sets based on peak traffic hours and extract the spatiotemporal effect of high mobility using incoming and outgoing handover relationships between cells. We perform multi-step prediction up to 24 hours ahead and propose a parametric SLA-based loss function to guarantee the SLA violation rate. We propose single-cell and multi-cell training architectures and assess their performance results. We predict slice-based traffic using a multi-slice architecture within a cell and calculate the total traffic at the cell level, achieving a lower SLA-based loss.

\section{Dataset and Additional Features}
\label{section:Datasetandcustomfeatures}

\begin{figure} [t]
	\centering
	\includegraphics[width=3.5in]{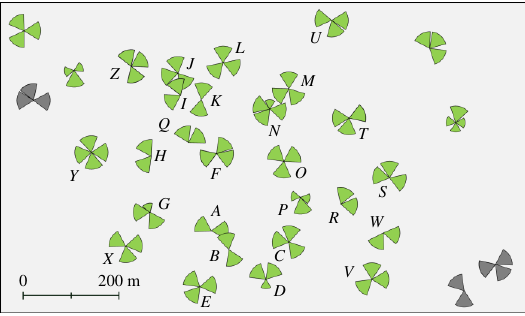}
	\caption{Location of cells in our dataset.}
	\label{fig:HighlyDenseUrbanArea}
\end{figure}

This section presents the dataset and methodology used in this study, which primarily focuses on predicting the downlink traffic volume in a live RAN that serves a densely populated urban area with high user mobility. The dataset encompasses 30 base stations, corresponding to 135 cells, due to the presence of multiple sectors and carriers within each base station (see Fig.~\ref{fig:HighlyDenseUrbanArea}). For this study, letters denote base stations, and numbers denote the cells within a given base station. For instance, $A_1$ represents the first cell of base station $A$, while $BS_n$ describes a generic cell within a generic base station.

\begin{figure} [t]
	\centering	\includegraphics[]{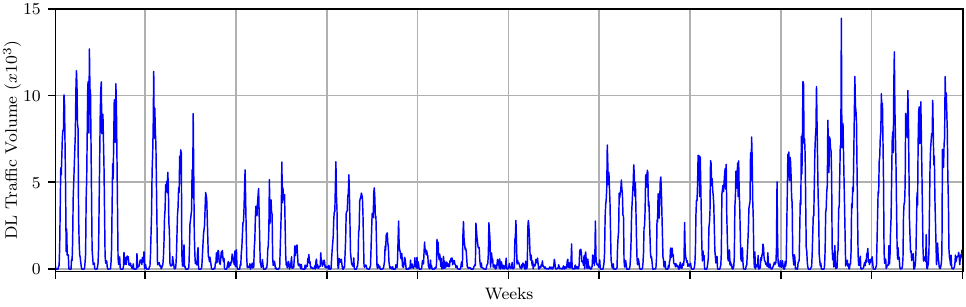}
	\caption{Total downlink traffic volume of cell ${A_2}$ for a 10-week period.}
	\label{fig:GU14yearlytraffic}
\end{figure}

The dataset comprises hourly measurements for each cell, encompassing both total and slice-based traffic metrics collected over a period of 52 weeks. Fig.~\ref{fig:GU14yearlytraffic} displays the hourly downlink traffic volume graph for a typical cell, $A_2$, located in a crowded city square. This cell experiences a heavy traffic load, with unexpected spikes during peak hours of the day. Besides daily fluctuations, the traffic volume is notably higher on weekdays than on weekends in some weeks. Moreover, the weekly average traffic volume changes week by week. These various dynamic behaviors make traffic prediction exceptionally challenging for such cells. We note that data aggregated at the base station or grid level granularity is significantly smoother due to the averaging of dynamic effects. The smoother the time series data, the better the prediction performance. Nevertheless, we choose to work with cell-level data, as it is essential for intelligent cell-level network control.

\begin{figure*}[t]
	\centering
	\subfloat[Voice slice.]{\includegraphics[]{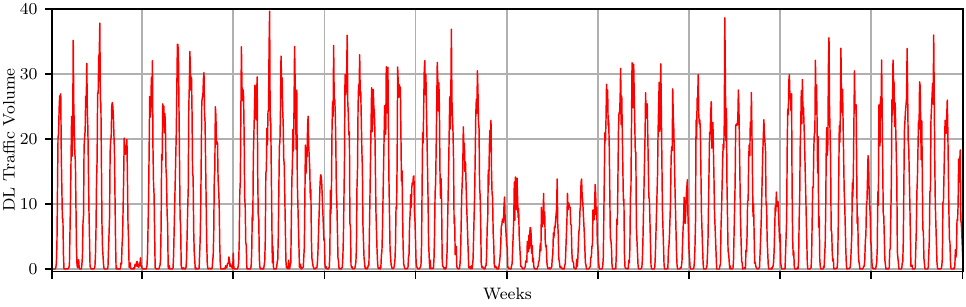}
		\label{fig:Plot_GU14_year_Voice}}
	\hfil
	\subfloat[Data slice.]{\includegraphics[]{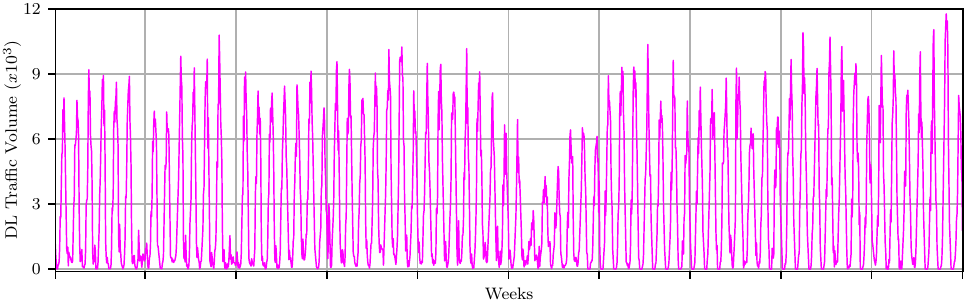}
		\label{fig:Plot_GU14_year_Dslice}}
	\hfil
	\subfloat[FWA slice.]{\includegraphics[]{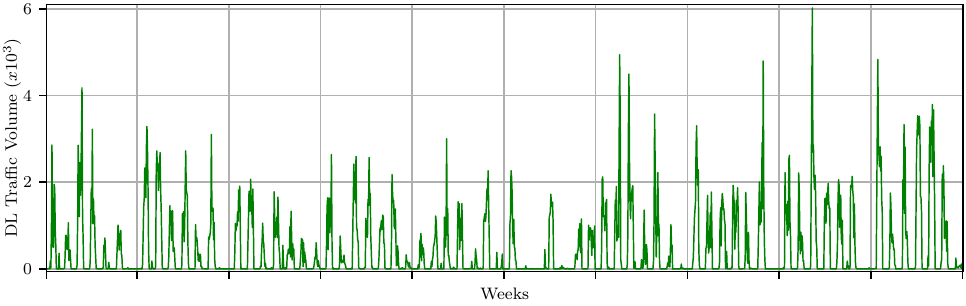}
		\label{fig:Plot_GU14_year_FWA}}
	\caption{Slice-based downlink traffic volume of cell ${A_2}$ for a 10-week period.}
	\label{fig:GU14_Slice_Year}
\end{figure*}

Slice-based measurements, available for a duration of 36 weeks, include traffic data for three distinct services: voice, data, and Fixed Wireless Access (FWA) for each cell. Fig.~\ref{fig:GU14_Slice_Year} illustrates the downlink traffic volume of each slice of cell $A_2$. The downlink traffic volume levels vary among each slice. Additionally, the voice and data slices display a more regular time series structure, while the FWA slice exhibits an irregular time series structure. These differences underscore the importance of monitoring slice-based traffic.

Table~\ref{table:Features_Abbreviations} outlines the collected features of mobile network traffic with their respective measurement family names \cite{3gpp.32.425, 3gpp.32.401}. The dataset includes 20 RAN features, with the total downlink traffic volume (labeled F0) being the output feature to be predicted.

\begin{table} [t]
\centering
\footnotesize
\caption{RAN Features}
\label{table:Features_Abbreviations}
\begin{tabular}{ll}
\toprule
\textbf{Label}	& {\textbf{Name}} 	\\ 
\midrule
F0  			& Downlink Traffic Volume                             \\  
F-RAN1  		& Avg. Num. of Active Users in downlink				\\  
F-RAN2  		& Avg. Num. of Active Users in uplink               \\ 
F-RAN3  		& Num. of Avg. Simultaneous RRC Connected Users \\ 
F-RAN4  		& Downlink PRB Utilisation                            \\ 
F-RAN5  		& Uplink Traffic Volume                             \\ 
F-RAN6  		& Uplink PRB Utilisation                            \\ 
F-RAN7  		& Num. of RRC Attempts                          \\ 
F-RAN8  		& Num. of S1 Signalling Establishment Attempt   \\
F-RAN9 		 	& Num. of Initial E-RABs Attempted to Setup     \\ 
F-RAN10 		& RACH Setup Succ. Rate                         \\ 
F-RAN11 		& Downlink PDCP Cell Thr.                             \\ 
F-RAN12 		& Downlink PDCP User Thr.                             \\
F-RAN13 		& Uplink PDCP Cell Thr.                             \\ 
F-RAN14 		& Uplink PDCP User Thr.                             \\ 
F-RAN15 		& Avg. Uplink RSSI Weight PUCCH                     \\ 
F-RAN16 		& Avg. Uplink RSRP PUSCH                            \\ 
F-RAN17 		& Avg. Uplink RSRP PUCCH                            \\ 
F-RAN18 		& Avg. RACH Timing Advance                      \\ 
F-RAN19 		& Avg. CQI                                      \\ 
\end{tabular}
\vspace{-12pt}
\end{table}

An objective of our research is to enhance the performance of our model by incorporating additional input features beyond the downlink traffic volume. Initially, we examine the correlation between the downlink traffic volume and the other monitored features in the RAN. We anticipate that incorporating highly correlated features into the input dataset will positively impact the model's performance. We also introduce an additional feature set named {\em Peak Hours Feature}, which emphasizes specific time periods during the day. Lastly, we propose the {\em Mobility Clustering} method to address the spatiotemporal effect, which factors in the incoming and outgoing handover relationships. Below, we elaborate on the methods we use to extract these features.

\subsection{RAN Features}
\label{subsection:subsubsection_RANFeaturesOnly}

\begin{figure}[t]
	\centering
	\includegraphics[]{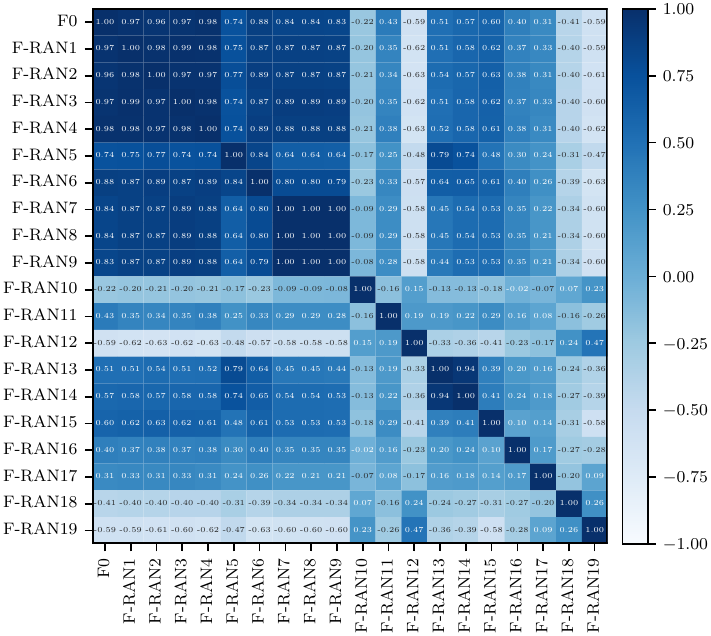}
	\caption{Correlation heatmap of RAN features for cell ${A_2}$.}
	\label{fig:GU14_Heatmap_Total}
\end{figure}

We consider 20 different RAN features, whose labels are provided in Table~\ref{table:Features_Abbreviations}. We compute the Pearson correlation coefficient between the downlink traffic volume feature (F0) and the other features to identify the most correlated features. Fig.~\ref{fig:GU14_Heatmap_Total} illustrates the correlation heatmap for cell $A_2$.

We establish a correlation threshold of 0.90 and examine the heatmap of various cells to determine the features that can be incorporated into the input dataset. In the total traffic dataset, we include features F-RAN1, F-RAN2, F-RAN3, and F-RAN4, in addition to the F0 feature. Conversely, for the slice-based traffic dataset, only F-RAN1 and F-RAN2 are included along with the F0 feature since F-RAN3 and F-RAN4 measurements are unavailable per slice. We denote the multivariate model that is based on RAN features as ``mvLSTM-RAN''.

\subsection{Peak Hours}
\label{subsection:subsubsection_PeakHours}

Our observations show that the F0 feature follows a 24-hour cycle for all cells, with intervals of low and high demand for traffic volume. In addition, F0 values are generally higher on weekdays and lower on weekends in densely populated urban areas. To capture these patterns, we introduce two additional Boolean feature vectors.

\begin{figure}[t]
	\centering
	\includegraphics[]{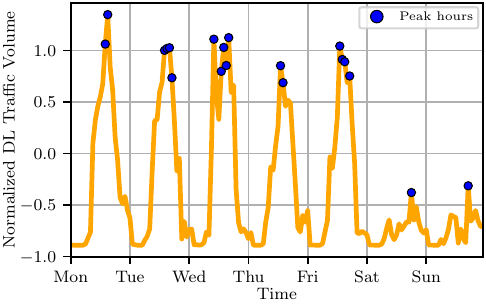}
	\caption{The labels for the peak hours of the day in cell $A_2$.}
	\label{fig:PHFlabelGU14}
\end{figure}

The first feature vector, ``days of the week," distinguishes between weekdays and weekends in the input dataset. The second feature vector, ``peak hours of the day," labels the hours that have traffic volumes above 70\% of the peak value. This differentiation helps distinguish peak hours from non-peak hours. Fig.~\ref{fig:PHFlabelGU14} showcases the labeling of the peak hours of the day for cell $A_2$, marked with circles.

We refer to this multivariate model as ``mvLSTM-peak'', which integrates the ``days of the week" and ``peak hours of the day" vectors in addition to the F0 feature. It is important to note that these additional features are derived from the statistics of F0 and do not necessitate further RAN measurements.

\subsection{Mobility Clustering}
\label{subsection:subsubsection_Handover}

\begin{figure}[t]
	\centering
	\includegraphics[width=3in]{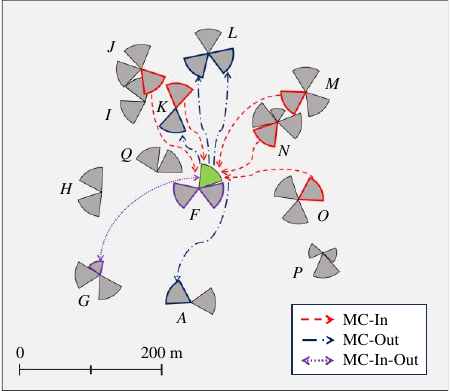}
	\caption{Mobility clustering for cell ${F_4}$.}
	\label{fig:HCclusterF17}
\end{figure}

In densely populated areas with highly mobile users, the traffic volume demand experiences dynamic changes as users transition between cells. To better understand, let us concentrate on the coverage area of a highly dynamic cell, $F_4$, which exhibits significant handovers. Fig.~\ref{fig:HCclusterF17} showcases our examination of the handover relationships of cell $F_4$, denoted in green. The arrows in Fig. \ref{fig:HCclusterF17}—represented by red dashed, blue dashed dot, and purple dot lines—indicate the incoming, outgoing, and bidirectional handover relations of the cell.

Multiple strategies can be used to leverage the interaction between cells. The first approach is proximity clustering, where the features of all cells in the region help predict the traffic volume of the target cell, $F_4$. When we include all cells within 200 meters of the target cell, the prediction encompasses all cells in base stations $A, F, G, H, I, J, K, L, M, N, O, P$, and $Q$, denoted in gray. When we include cells whose coverage area intersects with the target cell, the prediction involves all cells in base stations $K, L, N, O$, and $Q$. In both instances, the proximity clustering approach includes cells with no handover relationship to the target cell and excludes cells with a handover relationship. We assert that the handover relationship is the only means by which the traffic demand in one cell impacts the traffic demand in another cell. Therefore, we conclude that selecting cells in the cluster based solely on proximity and/or intersection of coverage does not improve prediction performance, underscoring the limitations of the grid structure of CNN-based approaches.

\begin{table}[t]
	\centering
	\footnotesize
	\caption{The rates of incoming and outgoing handovers for cell ${F_4}$}
	\label{tab:handoversy17}
	\begin{tabular}{@{}lc|lc@{}}
		\toprule
		\multicolumn{2}{c}{\textbf{Incoming}} & \multicolumn{2}{c}{\textbf{Outgoing}} \\ 
		\cmidrule(r){1-2}  \cmidrule(l){3-4} 
		\multicolumn{1}{l}{\textbf{Cells}} & \multicolumn{1}{l}{\textbf{Rate \%}} & \multicolumn{1}{l}{\textbf{Cells}} & \multicolumn{1}{l}{\textbf{Rate \%}} \\ 
		\midrule
		${N_2}$ 	& 18.34 & ${F_1}$ 	& 32.44 \\
		${F_3}$ 	& 17.72 & ${N_2}$ 	& 17.83 \\
		${F_2}$ 	& 9.84  & ${F_3}$ 	& 7.18 \\
		${M_3}$ 	& 7.49  & ${M_3}$ 	& 6.75  \\
		${F_1}$ 	& 5.51  & ${G_3}$ 	& 4.38  \\
		$F_{12}$	& 4.88  & $F_{12}$	& 4.35  \\
		${G_3}$ 	& 4.68  & ${O_5}$ 	& 3.36  \\
		${J_1}$ 	& 4.32  & ${J_1}$ 	& 3.34  \\
		${O_5}$ 	& 4.23  & ${L_5}$ 	& 2.28  \\ 
		\bottomrule
	\end{tabular}
	\vspace{-12pt}
\end{table}

We propose the mobility clustering method to take advantage of the interaction between cells with handover relations. The cluster in this method only includes cells that maintain a handover relationship with the target cell $F_4$. Table~\ref{tab:handoversy17} lists the neighboring cells that have incoming and outgoing handover relations with $F_4$. With mobility clustering, we expand the input dataset by incorporating two feature vectors. We construct these vectors by calculating the weighted averages of the downlink traffic volumes of the neighboring cells—one vector for the incoming and the other for the outgoing handover relationship. We refer to this multivariate model as ``mvLSTM-handover'', which includes these two additional feature vectors derived through mobility clustering.

\section{Neural Network Model with Parametric Loss Function}
\label{section:Neuralnetworkmodelwithcustomlossfunctions}

Our primary objective is to prevent SLA violations by developing a machine learning-based prediction method that considers multiple RAN features, unexpected traffic spikes during peak hours, and handovers between cells. To achieve this, we employ a multivariate LSTM architecture. While more complex models are available, we choose LSTM for several reasons: first, these more complex models cater to a grid-based approach that does not align with our focus on individual cells; second, the lengthy inference time of these complex models is impractical for the near-real-time RIC; third, our emphasis lies in mobility clustering, multivariate features, and an SLA-based loss function. This approach serves to strike a balance between computational feasibility and predictive accuracy, ensuring optimal network performance and resource allocation.

\begin{figure} [t]
	\centering
	\includegraphics[]{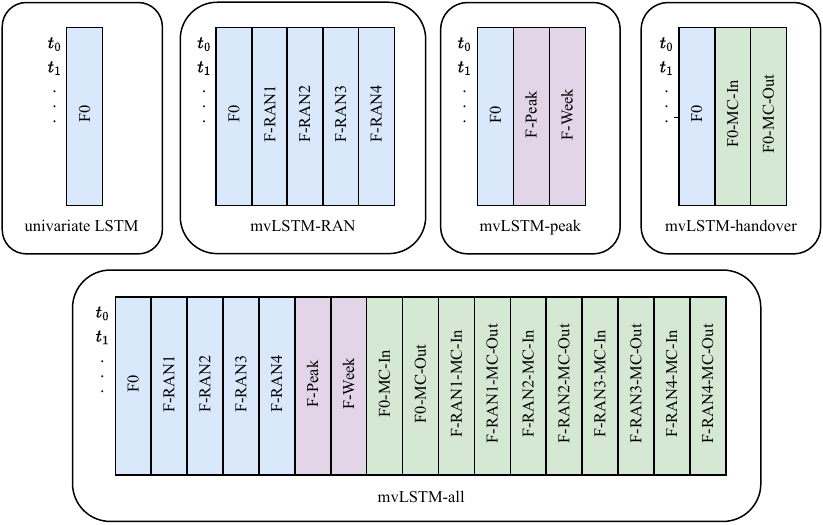}
	\caption{The input datasets of total traffic for different LSTM models.}
	\label{fig:LSTMmodels}
\end{figure}

We introduce four multivariate LSTM models for predicting traffic volume in the RAN. The first model, mvLSTM-RAN, integrates additional RAN features. The second model, mvLSTM-peak, concentrates on the peak hours of the day and days of the week. The third model, mvLSTM-handover, includes the weighted averages of downlink traffic volumes of cells within the mobility cluster. Lastly, the mvLSTM-all model integrates all three multivariate models. Fig.~\ref{fig:LSTMmodels} illustrates the total traffic input dataset structure of different LSTM models. We refer to the incoming and outgoing feature vectors acquired through the mobility clustering (MC) method as MC-In and MC-Out, respectively. For example, F0-MC-In signifies the vector obtained by weighted averaging the F0 vectors of the cells with incoming handover relationships to the relevant cell. It is worth noting that the mvLSTM-RAN model for slice-based traffic incorporates the F0, F-RAN1, and F-RAN2 features only.

To benchmark the performance of our multivariate model, we employ the univariate LSTM model as a reference. In the univariate model, the input comprises an array of the downlink traffic volume from the past 24 hours, and the output is the subsequent instance of the downlink traffic volume. Conversely, in the multivariate models, the input consists of a list of arrays, each containing the past 24 hours of a specific feature.

We utilize the K-fold cross-validation technique to accommodate changing traffic levels and trends while training the model. We split the total traffic dataset into training, validation, and test sets with durations of 40 weeks, 8 weeks, and 4 weeks, respectively. We use 6 folds, subdividing the 48-week dataset into 8-week segments. Conversely, we split the slice-based dataset into training, validation, and test sets with durations of 24 weeks, 8 weeks, and 4 weeks, respectively. We use 4 folds, breaking up the 32-week dataset into 8-week segments. Each execution yields different training and validation sets by shifting the training dataset by 8 weeks. We normalize the feature sets using the mean and standard deviation values calculated from the training set.

After preprocessing the data, we focus on the model's loss function. Standard loss functions such as MAE and MSE present two significant disadvantages: first, they impose identical penalties for both SLA violations and overprovisioning; second, they cannot be parameterized with an SLA violation rate. Consequently, we introduce a parametric SLA-based loss function that overcomes both disadvantages of standard loss functions.

Let us denote $y_n$ and $\mathbf{x}_n$ as the target value and input features of a prediction problem, respectively, where $n$ is the index of the elements in the dataset. Furthermore, let $f(\cdot)$ symbolize the input-output relationship of the DNN, determined by the network weights. We can express the predicted value as $\hat y_n(w) = f(\mathbf{x}_n, w)$, where $w$ represents a loss function parameter. The training process seeks to minimize the average of a loss function, $\mathcal{L}(\cdot, w)$, to determine the DNN weights by using
\begin{align}
	\min_f \frac{1}{N_{\text{train}}} \sum_{n=1}^{N_{\text{train}}}\mathcal{L}\left(f(\mathbf{x}_n, w) - y_n, w \right).
	\label{eqn:training}
\end{align} 

Let us denote the prediction error as $e_n(w) = \hat y_n(w) - y_n$. During test time, an SLA violation occurs when the prediction error is negative, $e_n(w) < 0$, and overprovisioning takes place when the prediction error is positive, $e_n(w) > 0$. We define the SLA violation rate as the percentage of instances when an SLA is violated
\begin{align}
	r(w) = \frac{1}{N_\text{test}} \sum_{n=1}^{N_\text{test}} \mathbbm{1}\left(-e_n(w)\right),
\end{align} 
where $N_\text{test}$ represents the number of elements in the test set, and $\mathbbm{1}(\cdot)$ serves as the indicator function. It is crucial to understand that the SLA violation rate is not synonymous with the uptime or the outage rate. When traffic demand is underprovisioned (i.e., an SLA violation occurs), the customer continues to receive service but at a slightly diminished rate. Operators commonly utilize SLA violation rates of 1\%, 3\%, 5\%, and 10\%.

We define the overprovisioning volume as the average positive prediction error
\begin{align}
	v(w) = \frac{1}{\sum_{t=1}^{N_\text{test}} \mathbbm{1}\left(e_n(w)\right)} \sum_{t=1}^{N_\text{test}} e_n(w) \mathbbm{1}\left(e_n(w)\right).
\end{align} 
After testing various parametric loss functions, we opt for the weighted Mean Absolute Error (wMAE) as the parametric loss function, which we define as
\begin{align}
\mathcal{L}_{\text{wMAE}}(\hat x - x, w) =  \left\{ 
\begin{array}{ll}
	w |\hat x - x|, & \quad e \leq 0 \\
	\hat x - x, & \quad  e > 0 \\
\end{array}.
\right.
\label{eqn:loss-wMAE}
\end{align}

Our goal is to minimize the SLA-based wMAE loss and overprovisioning volume while respecting a constraint on the SLA violation rate by choosing an appropriate $w$. We explore two different SLA violation rate scenarios, 3\% and 5\%, and conduct a line search to determine the optimal weights, denoted as $w_{3\%}$, and $w_{5\%}$. 

\section{Single-cell Prediction}
\label{section:Singlecellprediction}

\begin{figure} [t]
	\centering
	\vspace{0.1in}
	\includegraphics[]{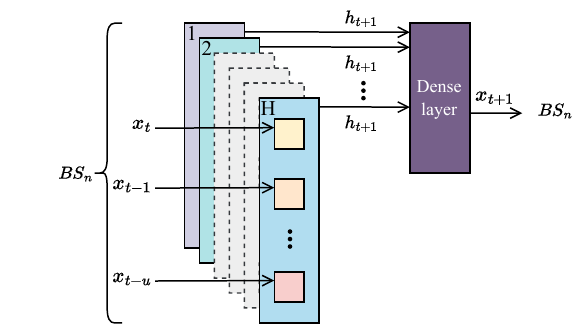}
	\caption{Single-cell architecture, where an LSTM layer with window size $u$ and $H$ hidden units is followed by a dense layer.}
	\label{fig:SingleCell}
\end{figure}

In this section, ``single-cell" refers to training a model using the data from a single cell. Fig.~\ref{fig:SingleCell} illustrates the single-cell architecture. The input features from a cell first undergo processing in the LSTM layer, where $H$ and $u$ represent the number of hidden units and window size, respectively. The output of the LSTM layer then moves into a dense layer, which produces F0 predictions for the cell in question. However, since a network might comprise thousands of cells, managing a separate model for each can become challenging and costly. We address this critical trade-off in subsequent sections. This section presents the single-step and multi-step prediction performance evaluation of the single-cell training architecture using the total traffic dataset.

\subsection{Single-step Prediction}
\label{section:section_Singlestep}

This subsection focuses on the effectiveness of our multivariate LSTM models in a single-step prediction scenario, where only the next time instance is predicted. We assess the performance of cell-based traffic volume predictions for typical base stations $A$ and $D$ under 3\% and 5\% SLA conditions. Specifically, we consider cells $A_1$, $A_2$, $A_3$, ${D_1}$, ${D_2}$, ${D_3}$, ${D_4}$, ${D_5}$, and ${D_6}$, all of which have different coverage areas.

\begin{table}[t]
	\caption{The comparison for the single-step prediction performance in cell $A_2$}
	\label{tbl:A14singlesteptraditional}
	\centering
	\footnotesize
	\begin{tabular}{ l|cc }
		\toprule
		\bf{Models}	& \bf{SLA-based loss} & \bf{SLA\%} \\
		\midrule 
		ARIMA              & 5.51 & 83.60 \\
		MSE-based LSTM     & 0.83 & 42.66 \\
		LogCosh-based LSTM & 0.67 & 31.55 \\
		Huber-based LSTM   & 0.66 & 30.56 \\
		MAE-based LSTM     & 0.63 & 29.56 \\
		Univariate LSTM    & 0.44 & $\leq$ 5 \\
		mvLSTM-RAN         & 0.43 & $\leq$ 5 \\
		mvLSTM-peak        & 0.44 & $\leq$ 5 \\
		mvLSTM-handover    & 0.39 & $\leq$ 5 \\
		mvLSTM-all         & 0.39 & $\leq$ 5 \\
		\bottomrule
	\end{tabular}
	\vspace{-12pt}
\end{table}

Before delving into the performance of our proposed LSTM models, it is essential to understand why univariate models using traditional loss functions are inadequate for our specific problem. Table~\ref{tbl:A14singlesteptraditional} presents the SLA-based loss values and corresponding SLA violation percentages of traditional models and our proposed models using a 5\% SLA violation constraint. Among the univariate LSTM models using traditional built-in loss functions, the most effective MAE-based LSTM incurs 43.2\% higher SLA-based loss than the univariate LSTM model. More importantly, our proposed models maintain a given SLA violation rate. These points highlight the importance of our proposed models for the network traffic prediction problem.

\begin{figure*}[t]
	\centering
	\subfloat[Predicted traffic for the MAE-based LSTM model]{\includegraphics[]{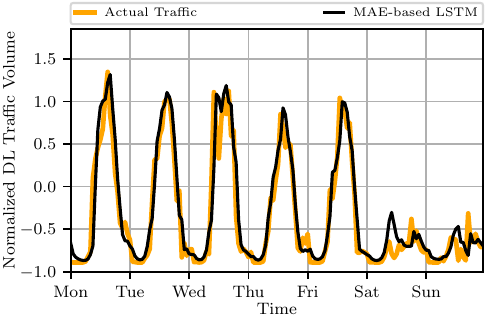}
		\label{fig:Plot_predact_MAEbased}}
	\subfloat[Predicted traffic for the  mvLSTM-peak model]{\includegraphics[]{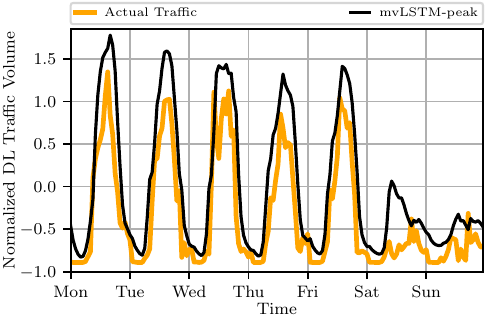}
		\label{fig:Plot_predactpeak}}
	\hfil
	\subfloat[Error bars for the MAE-based LSTM model]{\includegraphics[]{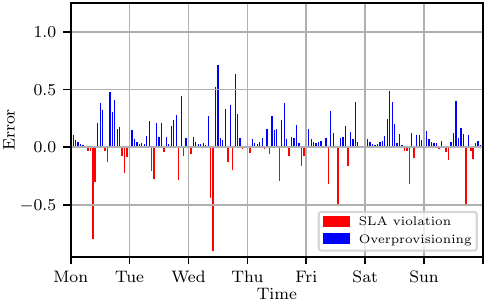}
		\label{fig:Plot_error_MAEbased}}
	\subfloat[Error bars for the mvLSTM-peak model]{\includegraphics[]{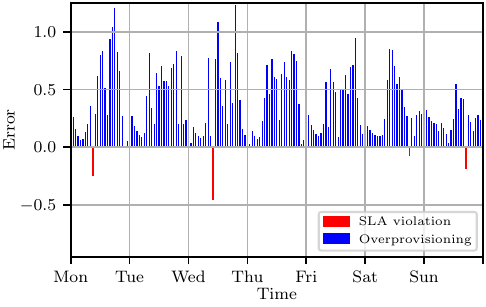}
		\label{fig:Plot_error_peak}}  
	\caption{Actual traffic, predicted traffic, and error values for cell $A_2$. Red lines are SLA violations.}
	\label{fig:A14_SLA5P_Actpred_MAE_peak}
\end{figure*}

Actual and predicted downlink traffic volumes appear in Fig.~\ref{fig:A14_SLA5P_Actpred_MAE_peak} for MAE-based univariate LSTM and mvLSTM-peak models. Although the predicted traffic curve in Fig.~\ref{fig:Plot_predact_MAEbased} seems to follow the actual traffic more closely than the predicted traffic curve in Fig.~\ref{fig:Plot_predactpeak}, error bars in Fig.~\ref{fig:Plot_error_MAEbased} show significant SLA violations for MAE-based LSTM. In contrast, Fig.~\ref{fig:Plot_error_peak} presents very few SLA violations. MAE-based LSTM models generally result in SLA violations about half the time and lag behind actual traffic during peak hours. For SLA-based LSTM models, it is possible to adjust the level of SLA violations by tuning the weight parameter.

Table~\ref{tab:benchmark-1step-GU} displays the cell-based test loss results for the $A_1$, $A_2$, and $A_3$ cells. SLA-based loss values for cell $A_1$ are much higher suggesting that traffic in $A_1$ is more dynamic. We observe that different models perform better for different cells. A change in the SLA violation rate also changes the best-performing model for $A_2$. When a lower SLA is targeted, using peak hours features proves more beneficial. These results underscore the limitations of a one-size-fits-all model for predicting multiple cells. Furthermore, they highlight the need for determining which additional features are most beneficial for individual cell traffic prediction.

\begin{table}[t]
	\centering
	\caption{SLA-based loss values for single-step prediction of base station $A$}
	\label{tab:benchmark-1step-GU}
	\footnotesize
	\begin{tabular}{@{}l|cccccc@{}}
		\toprule
		\multirow{2}{*}{\textbf{Models}} &
		\multicolumn{2}{c}{\textbf{$A_1$}} &
		\multicolumn{2}{c}{\textbf{$A_2$}} &
		\multicolumn{2}{c}{\textbf{$A_3$}} \\ \cmidrule(l){2-3}  \cmidrule(l){4-5}  \cmidrule(l){6-7}
		&
		\multicolumn{1}{c}{\textbf{3\%}} &
		\multicolumn{1}{c}{\textbf{5\%}} &
		\multicolumn{1}{c}{\textbf{3\%}} &
		\multicolumn{1}{c}{\textbf{5\%}} &
		\multicolumn{1}{c}{\textbf{3\%}} &
		\multicolumn{1}{c}{\textbf{5\%}} \\ \midrule
		univariate LSTM & 1.88      & 1.56      & 0.50      & 0.44      & 0.53      & 0.38 \\
		mvLSTM-RAN      & \bf{1.20} & \bf{1.09} & 0.49      & 0.43      & 0.64      & 0.48 \\
		mvLSTM-peak     & 1.43      & 1.23      & \bf{0.44} & 0.44      & 0.49      & 0.39 \\
		mvLSTM-handover & 1.70      & 1.17      & \bf{0.44} & \bf{0.39} & 0.63      & 0.52 \\
		mvLSTM-all      & 1.31      & 1.11      & 0.45      & \bf{0.39} & \bf{0.47} & \bf{0.38} \\ 
		\bottomrule
	\end{tabular}
	\vspace{-12pt}
\end{table}
\begin{figure}[t]
	\centering
	\includegraphics[width=5in]{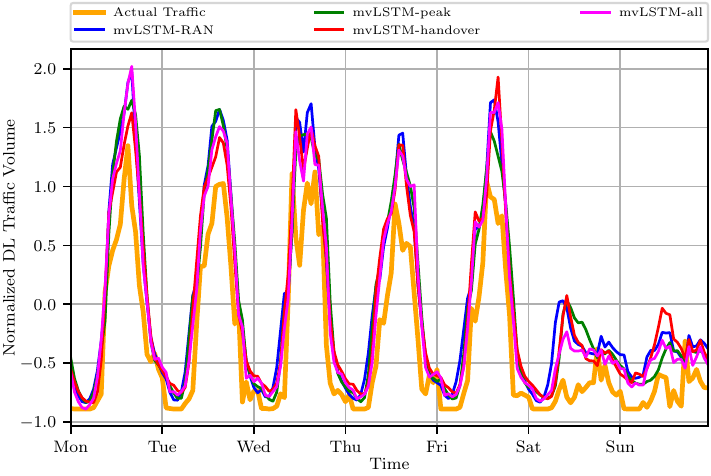}
	\caption{Actual and predicted values for cell $A_2$ under 3\% SLA violation rate.}
	\label{fig:GU14_Diff_SLAs_3P}
\end{figure}

Fig.~\ref{fig:GU14_Diff_SLAs_3P} illustrates the predicted traffic curves of all models for cell $A_2$ under a stringent 3\% SLA violation rate. Due to the strict SLA violation rate, nearly all models exhibit no negative errors. Nonetheless, the mvLSTM-peak model proves superior in this context, yielding the lowest SLA-based loss and overprovisioning.

Next, to validate our findings from base station $A$, we explore another typical base station. Table~\ref{tab:benchmark-1step-MS} illustrates the cell-based performance results for cells $D_1$ through $D_6$. For base station $D$, we again observe that different models perform better for different cells. Using the same model for all cells in the system is not optimal, given that cells from different regions exhibit entirely different trends. We conclude that operators should consider cell-based prediction performance when selecting a suitable prediction model for traffic volume prediction. As we have seen from the results in this subsection, even if different cells operate in the same base station, each has different coverage area characteristics and data patterns.

\begin{table}[t]
\centering
\caption{SLA-based loss values for single-step prediction of base station $D$}
\label{tab:benchmark-1step-MS}
\footnotesize
\begin{tabular}{@{}l|cccccccccccc@{}}
\toprule
\multirow{2}{*}{\textbf{Models}} &
  \multicolumn{2}{c}{\textbf{$D_1$}} &
  \multicolumn{2}{c}{\textbf{$D_2$}} &
  \multicolumn{2}{c}{\textbf{$D_3$}} &
  \multicolumn{2}{c}{\textbf{${D_4}$}} &
  \multicolumn{2}{c}{\textbf{${D_5}$}} &
  \multicolumn{2}{c}{\textbf{${D_6}$}}  \\ \cmidrule(l){2-3}  \cmidrule(l){4-5}  \cmidrule(l){6-7}  \cmidrule(l){8-9}  \cmidrule(l){10-11}  \cmidrule(l){12-13} &
  \multicolumn{1}{c}{\textbf{3\%}} &
  \multicolumn{1}{c}{\textbf{5\%}} &
  \multicolumn{1}{c}{\textbf{3\%}} &
  \multicolumn{1}{c}{\textbf{5\%}} &
  \multicolumn{1}{c}{\textbf{3\%}} &
  \multicolumn{1}{c}{\textbf{5\%}} &
  \multicolumn{1}{c}{\textbf{3\%}} &
  \multicolumn{1}{c}{\textbf{5\%}} &
  \multicolumn{1}{c}{\textbf{3\%}} &
  \multicolumn{1}{c}{\textbf{5\%}} &
  \multicolumn{1}{c}{\textbf{3\%}} &
  \multicolumn{1}{c}{\textbf{5\%}} \\ \midrule
  univariate LSTM & 0.61      & 0.53      & 0.79      & 0.78      & 0.78      & 0.70       & 1.52      & 1.40      & 0.45      & 0.39      & 0.75      & 0.66 \\
  mvLSTM-RAN      & 0.56      & 0.52      & 0.72      & 0.64      & 0.86      & 0.70       & 1.35      & 1.34      & 0.40      & 0.32      & 0.70      & 0.65 \\
  mvLSTM-peak     & \bf{0.51} & \bf{0.48} & 0.70      & 0.64      & 0.75      & 0.69       & 1.42      & 1.35      & 0.40      & 0.34      & \bf{0.67} & \bf{0.63} \\
  mvLSTM-handover & 0.60      & 0.54      & 0.73      & 0.71      & 0.84      & 0.72       & \bf{1.34} & \bf{1.28} & 0.39      & 0.32      & 0.73      & 0.72 \\
  mvLSTM-all      & 0.54      & 0.51      & \bf{0.66} & \bf{0.59} & \bf{0.75} & \bf{0.68}  & 1.36      & 1.40      & \bf{0.36} & \bf{0.29} & 0.68      & 0.66 \\ \bottomrule
\end{tabular}
\end{table}

Another observation is that a model's SLA-based loss decreases as the SLA percentage increases. The weight for a 3\% SLA is higher than for a 5\% SLA, causing the loss function to penalize SLA violation cases more severely under the 3\% SLA condition. Consequently, the model avoids violating the SLA more often, leading to increased test loss due to frequent overestimating of actual values.

\subsection{Multi-step Prediction}
\label{section:section_Multistep}

In this subsection, we evaluate the effectiveness of our multivariate LSTM models in a multi-step prediction scenario. This section examines how much performance is affected by using multi-step prediction at the cell level. Specifically, we perform 1-hour, 2-hour, 4-hour, 8-hour, and 24-hour ahead predictions for cells $A_2$ and $D_1$. In a multi-step prediction scenario, previously predicted values serve as valid past values for subsequent predictions. Therefore, the accuracy of the initial prediction is crucial in determining the performance of the subsequent predictions. 

\begin{figure*}[t]
	\centering
	\subfloat[Tue]{\includegraphics[]{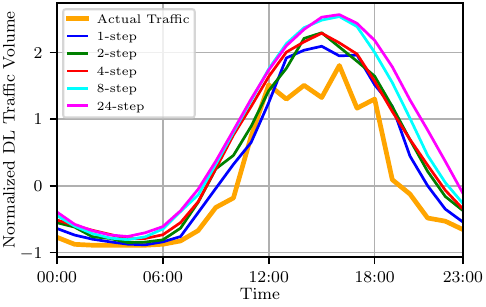}
		\label{fig:GU14_MS_Time1}}
	\subfloat[Thu]{\includegraphics[]{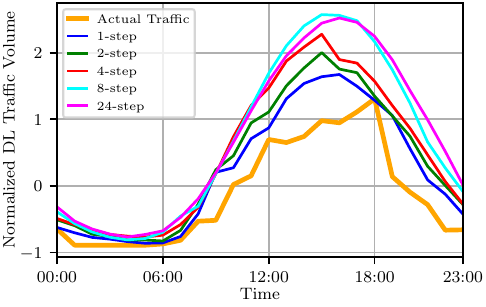}
		\label{fig:GU14_MS_Time2}}
	\caption{Multi-step prediction using the mvLSTM-handover model for cell $A_2$.}
	\label{fig:GU14_MS_Time}
\end{figure*}

Table~\ref{tab:benchmark-multistep-5} shows SLA-based loss values of multi-step prediction for cells $A_2$ and $D_1$. For the mvLSTM-handover model, the SLA-based loss for the 2-hour ahead prediction is 17.9\% higher than that of the 1-hour ahead prediction. As the number of steps in multi-step prediction increases, all models tend to overprovision more (higher loss values) during periods of rising and falling traffic demand. Furthermore, as the number of steps increases, the mvLSTM-peak model begins to outperform the mvLSTM-handover model. Fig.~\ref{fig:GU14_MS_Time} displays the delay in model predictions as the number of steps increases.

\begin{table}[t]
	\centering
	\caption{SLA-based loss values for multi-step prediction of cells $A_2$ and $D_1$ under 5\% SLA}
	\label{tab:benchmark-multistep-5}
	\footnotesize
	\begin{tabular}{@{}l|cccccccccc@{}}
		\toprule
		\multirow{2}{*}{\textbf{Models}} &
		\multicolumn{2}{c}{\textbf{1-step}} &
		\multicolumn{2}{c}{\textbf{2-step}} &
		\multicolumn{2}{c}{\textbf{4-step}} &
		\multicolumn{2}{c}{\textbf{8-step}} &
		\multicolumn{2}{c}{\textbf{24-step}} \\ \cmidrule(l){2-3}  \cmidrule(l){4-5}  \cmidrule(l){6-7}  \cmidrule(l){8-9}  \cmidrule(l){10-11}
		&
		\multicolumn{1}{c}{\textbf{$A_2$}} &
		\multicolumn{1}{c}{\textbf{$D_1$}} &
		\multicolumn{1}{c}{\textbf{$A_2$}} &
		\multicolumn{1}{c}{\textbf{$D_1$}} &
		\multicolumn{1}{c}{\textbf{$A_2$}} &
		\multicolumn{1}{c}{\textbf{$D_1$}} &
		\multicolumn{1}{c}{\textbf{$A_2$}} &
		\multicolumn{1}{c}{\textbf{$D_1$}} &
		\multicolumn{1}{c}{\textbf{$A_2$}} &
		\multicolumn{1}{c}{\textbf{$D_1$}} \\ \midrule
		univariate LSTM & 0.44      & 0.53      & 0.61      & 0.65      & 0.70      & 0.77      & 0.87      & 0.82      & 0.91      & 0.95 \\
		mvLSTM-RAN      & 0.43      & 0.52      & 0.49      & 0.60      & 0.56      & 0.62      & 0.77      & 0.79      & 1.00      & 1.08 \\
		mvLSTM-peak     & 0.44      & \bf{0.48} & 0.76      & 0.66      & \bf{0.50} & 0.60      & \bf{0.65} & \bf{0.69} & \bf{0.84} & \bf{0.89} \\
		mvLSTM-handover & \bf{0.39} & 0.54      & \bf{0.46} & \bf{0.57} & 0.57      & 0.63      & 0.89      & 0.83      & 1.07      & 0.91 \\
		mvLSTM-all      & \bf{0.39} & 0.51      & 0.50      & \bf{0.57} & 0.52      & \bf{0.58} & 0.70      & 0.74      & 0.87      & 0.98 \\ \bottomrule
	\end{tabular}
\end{table}

As the number of steps increases in multi-step prediction, models are more likely to experience SLA violations and overprovisioning, which can reduce performance. On the other hand, using multi-step prediction can provide computational advantages as long as the degradation in performance is tolerable. Multi-step prediction can be used in regions with more stable traffic demand changes.

\section{Multi-cell Prediction}
\label{section:Multicellprediction}

This section discusses the multi-cell training architecture for network traffic prediction. Multi-cell training employs a single model with a multi-input multi-output structure, which takes traffic from multiple cells belonging to either the same base station (single-BS) or different base stations (multi-BS) as input. As explained in Section~\ref{section:Singlecellprediction}, single-cell training can become cumbersome when the network contains numerous cells. However, the training time increases in multi-cell training due to having a much larger dataset size and model complexity. Most existing literature adopts the multi-BS approach. However, this section demonstrates that the performance of the multi-BS approach falls short of the single-cell approach. 

\begin{figure}[t]
	\centering
	\vspace{0.1in}
	\includegraphics[]{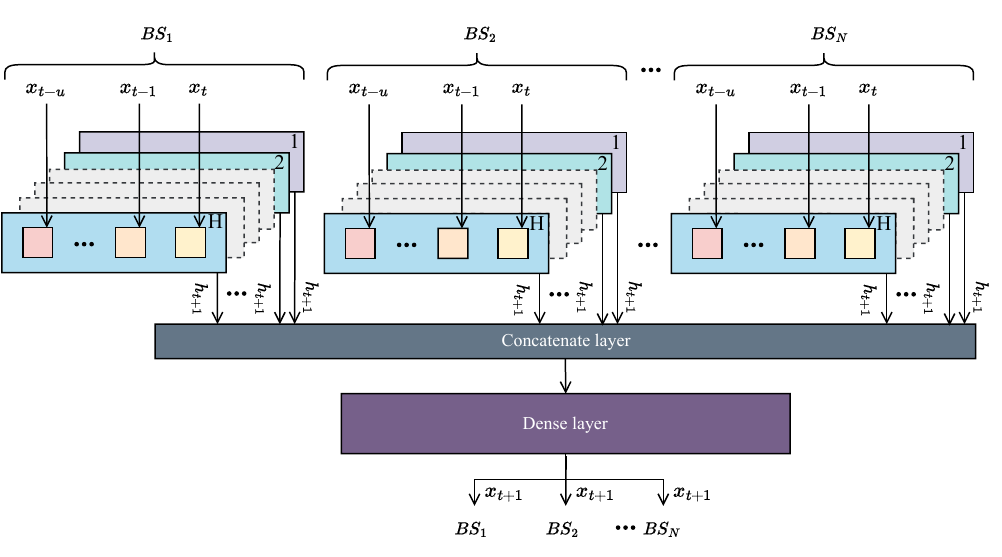}
	\caption{Multi-cell architecture.}
	\label{fig:MultiBS}
\end{figure}

The single-BS approach involves training a unified model architecture using traffic from all cells operated by a single base station. It's important to clarify that we do not aggregate cell-based traffic in this approach; instead, we predict the traffic of each cell separately. The scenario depicted in Fig.~\ref{fig:MultiBS} represents a single-BS architecture when the cells shown belong to the same base station.

On the other hand, multi-BS refers to training a larger model architecture using traffic from cells operating across multiple base stations. As with single-BS, no data aggregation is performed in the multi-BS approach, and we predict the traffic of each cell separately. The configuration shown in Fig.~\ref{fig:MultiBS} constitutes a multi-BS architecture when the cells indicated belong to different base stations.

We initially feed the features for each cell into an LSTM layer. We then concatenate the outputs of the LSTM layers. Finally, the concatenated outputs feed into separate dense layers to produce F0 predictions for each cell. Applying mobility-based clustering in these approaches is still possible since the handover information serves as an additional feature of a given cell.
\begin{table}[t]
	\centering
	\caption{SLA-based loss values for single-cell, single-BS and multi-BS prediction of cell $A_2$}
	\label{tab:benchmark-siso-mimo-m2mimo-GU14}
	\footnotesize
	\begin{tabular}{l|ccc}
		\toprule
		\textbf{Models} & \textbf{Single-cell} & \textbf{Single-BS} & \textbf{Multi-BS} \\ 
		\midrule
		univariate LSTM   & 0.44      & 0.44      & \bf{0.47}      \\
		mvLSTM-RAN        & 0.43      & 0.40      & \bf{0.47}      \\
		mvLSTM-peak       & 0.44      & 0.44      & 0.49           \\
		mvLSTM-handover   & \bf{0.39} & 0.38      & \bf{0.47}      \\
		mvLSTM-all        & \bf{0.39} & \bf{0.37} & 0.49           \\
		\bottomrule
	\end{tabular}
\end{table}  

Table~\ref{tab:benchmark-siso-mimo-m2mimo-GU14} exhibits the SLA-based loss values of single-cell, single-BS, and multi-BS predictions for cell $A_2$. The single-cell model considers the features of cell $A_2$ and generates that cell's prediction. Single-BS takes the features of the $A_1$, $A_2$, and $A_3$ cells from base station $A$ and provides predictions for all cells within $A$. In contrast, multi-BS takes in the features of all 135 cells in our dataset and outputs the predictions for all cells within this network. In Table~\ref{tab:benchmark-siso-mimo-m2mimo-GU14}, we only report the results of cell $A_2$. The mvLSTM-all model outperforms the rest, showing a 11.4\% and 15.9\% reduction in test loss for single-cell and single-BS prediction, respectively, compared to the univariate LSTM model. For multi-BS prediction, the convergence of test loss values across different models suggests that the effects of model-specific multivariate features are less pronounced. Notably, the performance degradation observed during the transition from single-cell to multi-BS, especially in the mvLSTM-all and mvLSTM-handover models, stresses the crucial role of handover effects for cell $A_2$. We can infer that complex and large models struggle to capture cell-specific information.

\begin{figure}[t]
	\centering
	\includegraphics[width=5in]{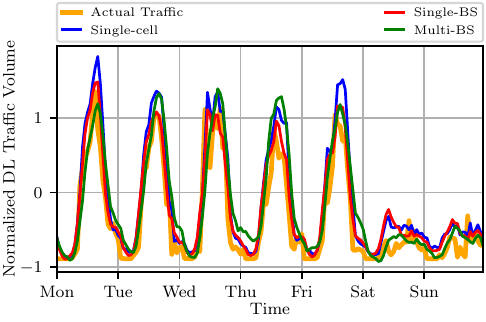}
	\caption{Single-cell, single-BS and multi-BS prediction using the mvLSTM-all model for cell $A_2$.}
	\label{fig:GU14_TRA_Time}
\end{figure}

Fig.~\ref{fig:GU14_TRA_Time} shows that both single-cell and single-BS prediction models tend to overprovision slightly more during peak traffic demand hours throughout the week compared to multi-BS predictions. This tendency is advantageous in avoiding SLA violations. Furthermore, the single-cell prediction curve is more adept at tracking traffic peaks than the other models. In contrast, the multi-BS prediction curve exhibits a smoother pattern, which unfavorably impacts its ability to prevent SLA violations.

\section{Slice-based Traffic Prediction}
\label{section:Slicebasedprediction}

In Sections~\ref{section:Singlecellprediction} and \ref{section:Multicellprediction}, we employed methods to predict the {\em total} downlink traffic volume. In this section, however, we broaden our approach to forecasting the downlink traffic volume for various services provided by network slices created within cells. Different services offered through network slices possess distinct traffic patterns and dynamics. These variations can often be overlooked when assessing total traffic volume. Therefore, to achieve slice-based traffic prediction, it is crucial to treat the traffic data of each service separately.

Our dataset includes three distinct services: voice, data, and FWA, each of which can function as a slice. Voice services have evolved into Voice over LTE (VoLTE) and Voice over 5G (Vo5G), offering high-quality voice calls coupled with video conferencing and multimedia messaging. With the advent of 5G, data services are set to reach unparalleled speed, capacity, and responsiveness, propelling investments in network infrastructure optimization. Additionally, FWA provides a flexible and cost-effective alternative to wired broadband connections, using wireless technology to deliver high-speed internet access to fixed locations without needing physical cables or fiber optics. FWA proves particularly valuable in challenging areas and is perfect for residential, business, and remote locations requiring reliable broadband connectivity.

This section presents single-slice and multi-slice methods for slice-based downlink traffic volume prediction in a single-cell scenario. The single-slice method trains a model using the dataset specific to a network slice within a cell. Conversely, the multi-slice method trains a model using the dataset comprising all slices within the same cell, similar to the single-BS approach in Section~\ref{section:Multicellprediction}.

Table~\ref{tab:benchmark-allslicessinglecellslicebased} illustrates the single-slice and multi-slice prediction performance results for cell $A_2$. Initially, we observe that the mvLSTM-peak and mvLSTM-all models perform best in the voice slice. Secondly, the mvLSTM-peak model exhibits the lowest SLA-based loss in the data slice in both scenarios. Particularly in the single-slice scenario, the gap in SLA-based loss between the mvLSTM-peak model and the univariate LSTM model reaches 25.5\%. The peak hours feature benefits voice and data slices due to their more regular time series structure. However, the FWA slice registers the highest test loss for the univariate LSTM compared to other slices, suggesting that using univariate features is inadequate to capture fluctuations in FWA service demand. The mvLSTM-all model, which performs best, yields an SLA-based loss 31.8\% lower than the univariate LSTM in single-slice prediction. Employing multivariate feature sets reduces the loss for all slices. Consequently, considering the demanding task of managing specific models for each slice, the multi-slice prediction method generally results in lower test loss than a single-slice prediction, making it the preferred approach.

\begin{table}[t]
\centering
\caption{SLA-based loss values for single-slice and multi-slice prediction of cell $A_2$}
\label{tab:benchmark-allslicessinglecellslicebased}
\footnotesize
\begin{tabular}{@{}l|cccccc@{}}
\toprule
\multirow{2}{*}{\textbf{Models}} &
  \multicolumn{2}{c}{\textbf{Voice}} &
  \multicolumn{2}{c}{\textbf{Data}} &
  \multicolumn{2}{c}{\textbf{FWA}} \\ \cmidrule(l){2-3} \cmidrule(l){4-5} \cmidrule(l){6-7}
  &
  \multicolumn{1}{c}{\textbf{Single-slice}} &
  \multicolumn{1}{c}{\textbf{Multi-slice}} &
  \multicolumn{1}{c}{\textbf{Single-slice}} &
  \multicolumn{1}{c}{\textbf{Multi-slice}} &
  \multicolumn{1}{c}{\textbf{Single-slice}} &
  \multicolumn{1}{c}{\textbf{Multi-slice}}
 \\ \midrule
  univariate LSTM & 0.40      & 0.34      & 0.51      & 0.37      & 0.88      & 0.61         \\
  mvLSTM-RAN      & 0.38      & 0.32      & 0.43      & 0.38      & 0.64      & 0.61         \\
  mvLSTM-peak     & \bf{0.34} & 0.33      & \bf{0.38} & \bf{0.34} & 0.72      & 0.61         \\
  mvLSTM-handover & 0.40      & 0.36      & 0.59      & 0.44      & 0.67      & 0.62         \\
  mvLSTM-all      & 0.36      & \bf{0.31} & 0.45      & 0.40      & \bf{0.60} & \bf{0.55}    \\ \bottomrule
\end{tabular}
\end{table}
\begin{figure*}[t]
	\centering
	\subfloat[Voice slice.]{\includegraphics[]{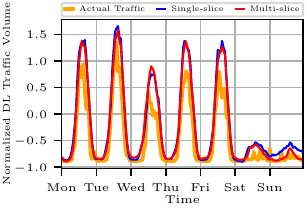}
		\label{fig:Plot_GU14_time_Voice}}
    \hfil
	\subfloat[Data slice.]{\includegraphics[]{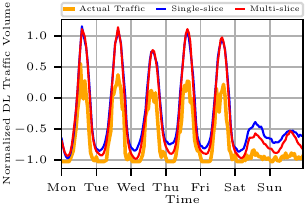}
		\label{fig:Plot_GU14_time_Dslice}}
    \hfil
	\subfloat[FWA slice.]{\includegraphics[]{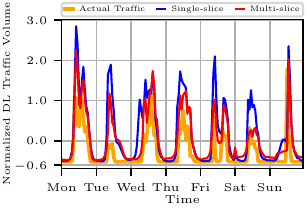}
		\label{fig:Plot_GU14_time_FWA}}
	\caption{Actual traffic and predicted traffic of single-slice and multi-slice prediction using the mvLSTM-all model for cell $A_2$.}
	\label{fig:GU14_Slice_Time}
\end{figure*}

Fig.~\ref{fig:GU14_Slice_Time} displays the actual and predicted traffic of single-slice and multi-slice prediction using the mvLSTM-all model for cell $A_2$. For the voice slice, the prediction values of the single-slice approach are generally higher than those of the multi-slice approach during peak hours and weekends, resulting in higher SLA-based loss for the single-slice method. The single-slice approach leads to more overprovisioning during non-peak hours and weekends in the data slice. The single-slice approach shows extremely high prediction values during peak traffic hours in the FWA slice. Additionally, on Wednesday, it causes significant overprovisioning as it accurately anticipates the surge in traffic demand a few hours in advance.

In addition to resulting in lower test loss and overprovisioning, another advantage of slice-based prediction is the ability to assign different SLA constraints to different services, which is the norm in NextG mobile networks. Table~\ref{tab:slicetuner} presents a comparison of test loss results at the slice level under different SLA conditions for total downlink traffic volume of cell $A_2$, using the mvLSTM-all model. Here, we compare the average of all slices under various SLA conditions with the single-cell result at 5\% SLA condition, as shown in Table~\ref{tab:benchmark-1step-GU}. The comparison reveals that as we assign different SLA conditions to different slices, the performance of slice-based prediction surpasses that of the single-cell prediction. Let us consider the configuration where the SLA constraints of voice, data, and FWA slices are 5\%, 10\%, and 15\%, respectively. Such a configuration prioritizes the voice slice, allowing for more SLA violations in the FWA slice. Compared to the mvLSTM-model and univariate LSTM model results listed in Table~\ref{tab:benchmark-1step-GU}, the slice-based prediction improvements in this configuration can rise to 28.2\% and 36.4\%, respectively.
\begin{table}[t]
\centering
\footnotesize
\caption{SLA-based loss values for multi-slice prediction with slice specific SLA constraints}
\label{tab:slicetuner}
\begin{tabular}{@{}c|ccc|c|c@{}}
\toprule
{\textbf{SLA\%}} &
{\textbf{Voice}} &
{\textbf{Data}} &
{\textbf{FWA}} &
{\textbf{Average}} & 
{\textbf{No slice}} \\ 
\midrule
    (1, 5, 10)  & 0.42 & 0.32 & 0.43 & 0.39 & 0.39 \\
    (3, 5, 10)  & 0.34 & 0.30 & 0.39 & 0.34 & 0.39 \\
    (5, 10, 15) & 0.35 & 0.23 & 0.24 & 0.28 & 0.39 \\ \bottomrule
\end{tabular}
\vspace{-12pt}
\end{table}
\section{Conclusion}
\label{section:section_Conclusion}

In this study, we have delved into the challenging problem of traffic prediction for NextG mobile networks, emphasizing the need for intelligent management and optimization of these complex systems. Our method incorporates 20 RAN features, additional features based on peak traffic hours, and a unique mobility-based clustering scheme to capture and utilize spatiotemporal effects. The devised parametric loss function is crucial in safeguarding the SLAs violation rate while reducing overprovisioning and ensuring reliable service delivery and resource utilization.

The paper thoroughly examined single-cell, multi-cell, and slice-based prediction approaches. Each approach brings unique strengths and challenges, shedding light on the intricacies of managing and predicting traffic in extensive mobile network environments. The exploration of slice-based traffic prediction is particularly relevant with the rise of network slicing in 5G networks. By focusing on the downlink traffic volume of various services offered within network slices, we pave the way for precise resource allocation and optimization.

Our comparative analysis highlights the performance and accuracy of the multi-slice prediction approach. This result signifies the potential of this method in effectively handling the diverse and dynamically changing traffic patterns within NextG networks. In addition, we find that using the same model for all cells in the network is not optimal, given that cells from different regions exhibit entirely different trends. 

However, the scope for further research and improvement remains. As NextG networks evolve, the complexity and dynamism of traffic patterns will continue to increase, necessitating even more refined prediction techniques. Future research can explore transferring the single-cell approach to an extensive network, collecting more mobility data, or developing an adaptive loss function to adjust to the changing SLA requirements.

\bibliographystyle{IEEEtran}

\end{document}